\documentclass[twocolumn]{aastex62}

\usepackage{graphicx}	
\usepackage{amsmath}	
\usepackage{amssymb}	
\usepackage[para,flushleft]{threeparttable} 
\usepackage{epstopdf}
\usepackage{textgreek}
\usepackage{comment}

\usepackage{xcolor, fontawesome}
\definecolor{twitterblue}{RGB}{64,153,255}
\newcommand{\twitter}[1]{\href{https://twitter.com/#1}{\textcolor{twitterblue}{\faTwitter}\,\tt\hspace{2pt}\textcolor{blue!60!black}{@#1}}}

\hypersetup{colorlinks,linkcolor={red!60!black},citecolor={blue!60!black},urlcolor={blue!60!black}}

\usepackage{natbib,twoopt}
\bibpunct{(}{)}{;}{a}{}{,} 
\makeatletter
\newcommandtwoopt{\citeads}[3][][]{\href{http://ui.adsabs.harvard.edu/\#abs/#3}%
{\def\hyper@linkstart##1##2{}%
\let\hyper@linkend\@empty\citealp[#1][#2]{#3}}}
\newcommandtwoopt{\citepads}[3][][]{\href{http://ui.adsabs.harvard.edu/\#abs/#3}%
{\def\hyper@linkstart##1##2{}%
\let\hyper@linkend\@empty\citep[#1][#2]{#3}}}
\newcommandtwoopt{\citetads}[3][][]{\href{http://ui.adsabs.harvard.edu/\#abs/#3}%
{\def\hyper@linkstart##1##2{}%
\let\hyper@linkend\@empty\citet[#1][#2]{#3}}}
\newcommandtwoopt{\citeyearads}[3][][]%
{\href{http://ui.adsabs.harvard.edu/\#abs/#3}
{\def\hyper@linkstart##1##2{}%
\let\hyper@linkend\@empty\citeyear[#1][#2]{#3}}}
\makeatother

\received{July 22, 2019}
\revised{August 29, 2019}
\accepted{\today}
\submitjournal{ApJL}

\shorttitle{Coronal response to magnetically-suppressed CME events in M-dwarf stars}
\shortauthors{Alvarado-G\'omez et al.}

\begin{document}

\title{\sc Coronal response to magnetically-suppressed CME events in M-dwarf stars}

\correspondingauthor{Juli\'an D. Alvarado G\'omez\\ \twitter{AstroRaikoh} }
\email{jalvarad@cfa.harvard.edu}

\author[0000-0001-5052-3473]{Juli\'an~D.~Alvarado-G\'omez}
\author[0000-0002-0210-2276]{Jeremy~J.~Drake}
\author[0000-0002-2470-2109]{Sofia~P.~Moschou}
\affil{Center for Astrophysics $|$ Harvard \& Smithsonian, 60 Garden Street, Cambridge, MA 02138, USA}

\author[0000-0002-8791-6286]{Cecilia Garraffo}
\affiliation{Institute for Applied Computational Science, Harvard University, Cambridge, MA 02138, USA}

\author[0000-0003-3721-0215]{Ofer~Cohen}
\affil{University of Massachusetts at Lowell, Department of Physics \& Applied Physics, 600 Suffolk Street, Lowell, MA 01854, USA}
\collaboration{(NASA LWS Focus Science Team: The Solar-Stellar Connection)}

\author[0000-0002-9569-2438]{Rakesh K. Yadav}
\affiliation{Department of Earth and Planetary Sciences, Harvard University, Cambridge, MA 02138, USA}

\author[0000-0002-5456-4771]{Federico Fraschetti}
\affil{Center for Astrophysics $|$ Harvard \& Smithsonian, 60 Garden Street, Cambridge, MA 02138, USA}
\affil{Dept. of Planetary Sciences-Lunar and Planetary Laboratory, University of Arizona, Tucson, AZ, 85721, USA}



\begin{abstract}

\noindent We report the results of the first state-of-the-art numerical simulations of Coronal Mass Ejections (CMEs) taking place in realistic magnetic field configurations of moderately active M-dwarf stars. Our analysis indicates that a clear, novel, and observable, coronal response is generated due to the collapse of the eruption and its eventual release into the stellar wind. Escaping CME events, weakly suppressed by the large-scale field, induce a flare-like signature in the emission from coronal material at different temperatures due to compression and associated heating. Such flare-like profiles display a distinctive temporal evolution in their Doppler shift signal (from red to blue), as the eruption first collapses towards the star and then perturbs the ambient magnetized plasma on its way outwards. For stellar fields providing partial confinement, CME fragmentation takes place, leading to rise and fall flow patterns which resemble the solar coronal rain cycle. In strongly suppressed events, the response is better described as a gradual brightening, in which the failed CME is deposited in the form of a \textit{coronal rain cloud} leading to a much slower rise in the ambient high-energy flux by relatively small factors ($\sim2-3$). In all the considered cases (escaping/confined) a fractional decrease in the emission from mid-range coronal temperature plasma occurs, similar to the coronal dimming events observed on the Sun. Detection of the observational signatures of these CME-induced features requires a sensitive next generation X-ray space telescope.

\end{abstract}



\section{Introduction}\label{sec:intro}

\noindent Of the different manifestations of solar/stellar magnetic activity, flares and Coronal Mass Ejections (CMEs) are the most dramatic in terms of energetics and temporal evolution. It is widely accepted that they are interlinked phenomena, sharing a common origin in the energy transformation from magnetic into kinetic (particles and plasma) and radiative power (\citeads{2011LRSP....8....1C}, \citeads{2012LRSP....9....3W}, \citeads{2017LRSP...14....2B}). These transient events are expected to play an important role in the evolution of stellar rotation and activity (\citeads{2013ApJ...764..170D}, \citeads{2017ApJ...840..114C}), as well as in the conditions for retention of exoplanet atmospheres and habitability (\citeads{2018haex.bookE..19M}, \citeads{2019AsBio..19...64T}). Most of our knowledge of their origins, properties, and evolution comes from observations of the Sun, which indicate that large solar flares are typically accompanied by a CME (\citeads{2009EM&P..104..295G}, \citeads{2017SoPh..292....5C}).    

While strong flares are routinely detected in active stars, particularly M-dwarfs (see \citeads{2019ApJ...876..115S} and references therein), recent studies have shown that the solar flare-CME paradigm might be different in the stellar regime, such that the occurrence rate of CMEs and their associated kinetic energies are reduced dramatically (\citeads{2019A&A...623A..49V}, \citeads{2019ApJ...877..105M}). The recent direct detection of a stellar CME by \citetads{2019NatAs.tmp..328A} follows the same trend. In \citetads{2018ApJ...862...93A} we showed that these properties are expected due to the interaction between the escaping eruption and the suppressing effect imposed by the large-scale magnetic fields present in these stars. 

The growing interest in close-in, habitable zone planets orbiting M-dwarfs motivates a better understanding of the CME magnetic confinement process, especially given the relatively strong magnetic fields reported for these stars (e.g., \citeads{2010MNRAS.407.2269M}, \citeads{2017NatAs...1E.184S}). This letter contains the results of the first three-dimensional numerical simulations of CMEs developing within surface magnetic field values representative of moderately active M-dwarf stars. Our state-of-the-art models explore the response of the high-energy stellar corona to the different regimes of the CME confinement spectrum, analyzing specific observables in the context of previous solar and stellar studies, and their possible detection with future instrumentation.

\section{Numerical Models}\label{sec:models}

\noindent We proceed in a similar manner to \citetads{2018ApJ...862...93A}, first obtaining a steady-state description of the corona and stellar wind, and then using this solution as an initial condition of a time-dependent flux-rope CME simulation. The models employed here are incorporated in the latest version of the \href{http://csem.engin.umich.edu/tools/swmf/}{Space Weather Modeling Framework} (SWMF, see \citeads{2018LRSP...15....4G}).     

\subsection{Corona and Stellar Wind}\label{sec:ss}

\noindent Our M-dwarf corona and stellar wind simulations are constructed with the aid of the Alfv\'en Wave Solar Model (AWSoM, \citeads{2014ApJ...782...81V}), and the 3D MHD code BATS-R-US \citepads{2012JCoPh.231..870T}. This model, originally developed for solar system studies and adapted for the stellar regime, considers the distribution of the radial magnetic field on the surface of the star as a boundary condition. The field strength and polarity are used to construct an Alfv\'en wave turbulent dissipation spectrum that provides self-consistent heating of the corona and acceleration of the stellar wind. These Alfv\'en wave-driven contributions are incorporated as sources in the energy and momentum equations which, combined with the magnetic induction and mass conservation equations, complete the set of non-ideal MHD equations solved by the code. Effects of  radiative losses and electron heat conduction are also included in the simulation (see \citeads{2014ApJ...782...81V} for additional details). The numerical scheme evolves until a steady-state solution is obtained in the domain, described by a spherical grid extending from $\sim$\,$1.0$ $R_{*}$ to $85$ $R_{*}$ for the simulations discussed in this work. 
 
\subsubsection{Stellar Magnetic Field}\label{sec:field}

\noindent We have previously used surface magnetic field maps reconstructed with the technique of Zeeman-Doppler Imaging (ZDI, \citeads{1997MNRAS.291..658D}, \citeads{2002A&A...388..868K}) to simulate the corona and stellar wind environment around specific systems (Alvarado-G\'omez~et~al.~\citeyearads{2016A&A...588A..28A}, \citeyearads{2016A&A...594A..95A}, \citeads{2018ApJ...856...53P}), or as magnetic proxies for objects without observations (\citeads{2017ApJ...843L..33G}, \citeads{2019ApJ...875L..12A}).
Unfortunately, ZDI is not sensitive to the small-scale field expected to power CME activity and instead we drive models using the field topology, down to active region length-scales, resulting from a self-consistent dynamo simulation of a slowly-rotating fully-convective star (\citeads{2016ApJ...833L..28Y}). This dynamo model not only has been tailored to match the mass, radius, and rotation period of Proxima Centauri (hereafter, Prox Cen), but also yields a long-term cyclic magnetic field evolution that roughly captures the observed time-scale of the activity cycle in this star ($P_{\rm cyc} \sim 7$ years, see \citeads{2016A&A...595A..12S}, \citeads{2017MNRAS.464.3281W}). \citetads{2017ApJ...834...14C} followed a similar approach to study the coronal structure of rapidly-rotating M-dwarf stars.      
 
To ensure that the simulated magnetic field geometry is a robust representation of Prox Cen and other moderately active M-dwarfs, the field distribution is extracted at a time in which the dynamo solution is well within a cyclic regime\footnote[1]{This corresponds to approximately 490 rotations in the simulation performed by \citetads{2016ApJ...833L..28Y}.}. As described by \citetads{2016ApJ...833L..28Y}, the outer boundary for the dynamo simulation was set at $95\%$ of the stellar radius, which for the purposes of this study will be considered as the effective photosphere of the star. We apply five different scalings (preserving the field topology) to adjust the maximum surface radial field strength, ranging from $\pm600$~G to $\pm1400$~G in $200$~G increments. These values are consistent with magnetic field measurements in low to medium activity M-dwarf stars (\citeads{2014IAUS..302..156R}). Note that the $B_{\rm r,\, max}^{\rm range} = \pm1400$~G case has a mean surface magnetic field of $\left<B_{\,}\right>_{\rm s} \sim 450$~G, compatible with the low end estimate for Prox Cen reported from Zeeman broadening ($600 \pm 150$~G, \citeads{2008A&A...489L..45R}). All the models presented here consider currently accepted values for the properties of this star ($M_{*} = 0.122$~M$_{\odot}$, $R_{*} = 0.154$~R$_{\odot}$, $P_{\rm rot} = 83.0$ days; \citeads{2007AcA....57..149K}, \citeads{2017A&A...598L...7K}).  

\subsection{Flux-Rope Eruption}\label{sec:cme}

\noindent To initialize our CME simulations, we consider the analytical \citetads[][TD]{1999A&A...351..707T} flux-rope eruption model. This model is based on the magnetic pressure/tension imbalance acting on a twisted arc-like structure (sustained by a force-free field), which is embedded in the background magnetic field obtained in the steady-state stellar wind solution. The numerical implementation used here has been widely applied in detailed comparative studies in the solar context (e.g.~\citeads{2008ApJ...684.1448M}, \citeads{2013ApJ...773...50J}).

\begin{figure}[!t]
\centering
\includegraphics[trim=0.0cm 0.1cm 0.1cm 0.1cm, clip=true,width=0.478\textwidth]{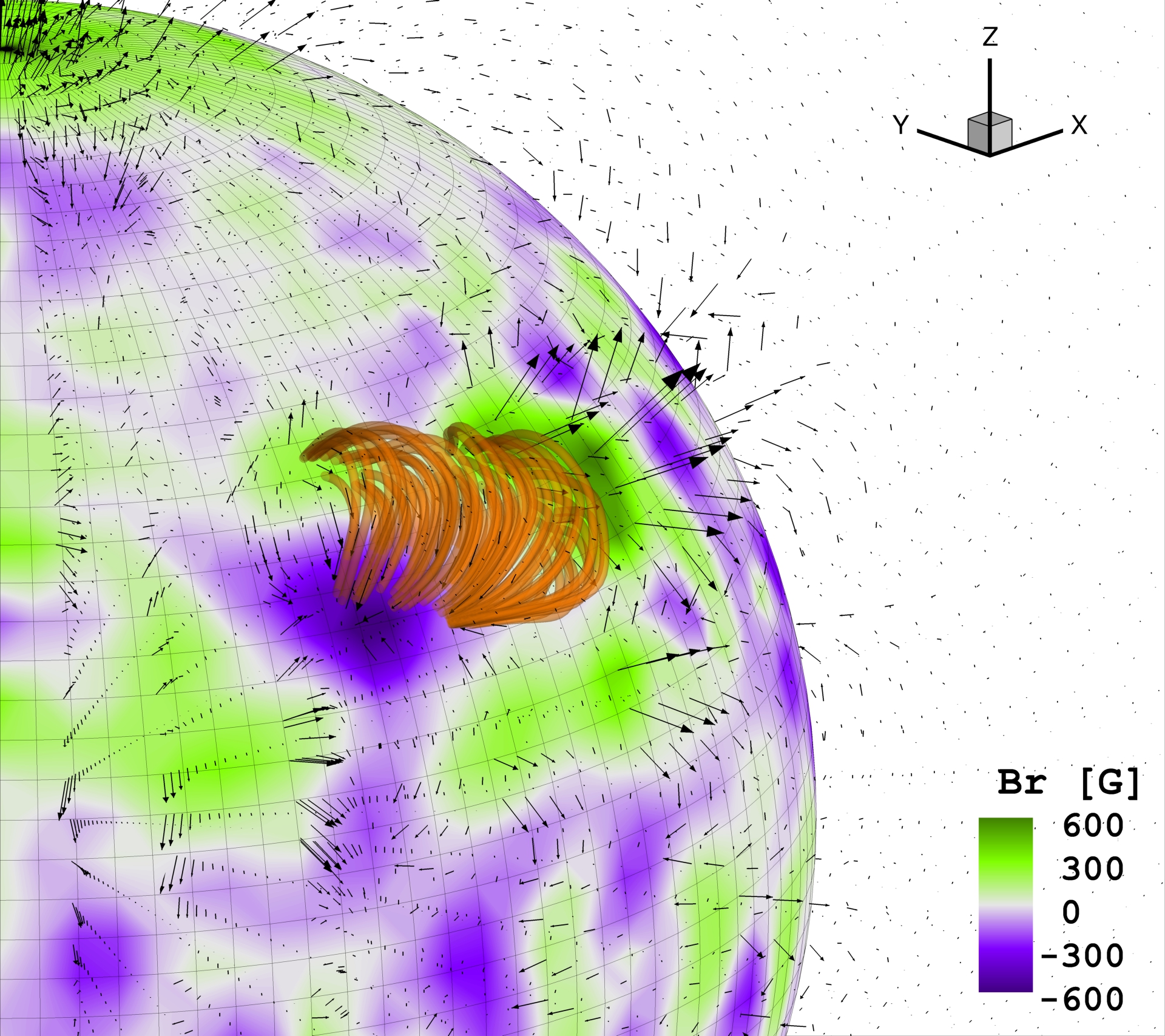}
\caption{\small Initial configuration of the CME simulation. The stellar surface is color-coded by the radial component of the magnetic field ($B_{\rm r}$), predicted by the fully-convective dynamo model of \citetads{2016ApJ...833L..28Y}. The Titov-D\'emoulin (TD) flux-rope (orange) is anchored to the strongest mixed-polarity region on the surface, as indicated by the black field vectors. The visualization shows the $B_{\rm r,\,max}^{\rm range} = \pm600$ G case.}
\label{fig_1}
\end{figure}

Eight parameters are required to determine the starting state of the TD model, including the position, orientation, and geometrical properties of the flux-tube, together with the electric current (used to compute the flux-rope magnetic free energy, $E_{\rm B}^{\rm FR}$) and mass loaded into the structure. As illustrated in Fig.~\ref{fig_1}, we anchor and align the flux-rope foot points to the strongest mixed-polarity region on the stellar surface, fixing in this way the location (longitude: $270$~deg, latitude: $36$~deg, orientation: $28$~deg), and size of the TD flux-tube (length:~$\sim$\,$150$~Mm, radius: $20$ Mm). 

The eruption is initialized with a magnetic free energy $E_{\rm B}^{\rm FR} \simeq 6.57 \times 10^{34}$ ergs (associated with an electric current $I_{0} = 8.0 \times 10^{12}$ A), carrying a mass of $M^{\rm FR}~=~4.0 \times 10^{14}$~g. Our selection of $E_{\rm B}^{\rm FR}$ lies slightly above the high-end values observed in eruptive solar active regions and CMEs (\citeads{2009EM&P..104..295G}, \citeads{2012LRSP....9....3W}, \citeads{2017ApJ...834...56T}), and should be easily achievable during typical flaring---and presumably CME---events in Prox Cen and other M-dwarf stars (cf.~\citeads{2015ApJ...809...79O}, \citeads{2018ApJ...860L..30H}). A nominal solar value is assumed for $M^{\rm FR}$ (\citeads{2011LRSP....8....1C}), as the final mass sweep off in the CME evolution is expected to be larger (by~$\sim$\,$1-2$ orders of magnitude), particularly due to the relatively denser corona and stellar wind compared to typical solar conditions.  

For each surface magnetic field scaling (see~Sect.~\ref{sec:field}), we follow the evolution of the eruption for 1.5 hours of real time, which is sufficient to cover the travel time of a very slow ($\sim25$ km s$^{-1}$) coronal perturbation through the inner corona (up to $R \geq 2.0$~$R_{*}$). We extract the MHD properties in the entire 3D domain at a cadence of 1 minute. 

\section{Results and Discussion}\label{sec:results}

\begin{figure*}[!t]
\centering
\includegraphics[trim=0.1cm 0.1cm 0.1cm 0.1cm, clip=true,width=0.995\textwidth]{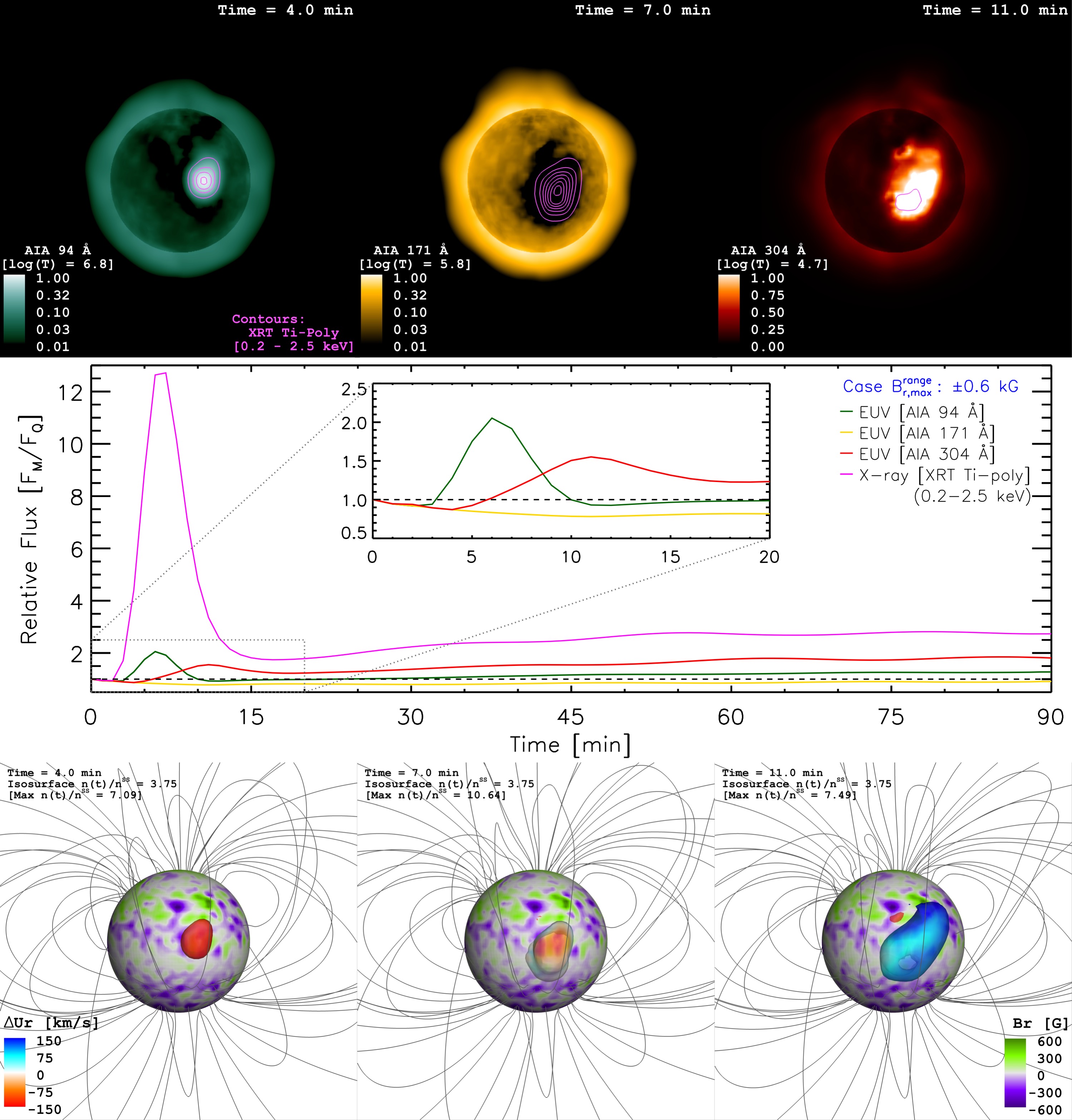}
\caption{Coronal response due to the flux-rope eruption in the $B_{\rm r,\,max}^{\rm range} = \pm600$ G case. \textit{Top:} Simulated line-of-sight images of the stellar corona (in arbitrary normalized units), synthesized in three different AIA/SDO filters (\textit{left:} 94 \AA; \textit{center:} 171~\AA; \textit{right:} 304~\AA) at different times during the evolution of the CME. Contours in magenta show the location of the resulting X-ray emission (Ti-poly filter of XRT/Hinode; $0.2-2.5$ keV), spaced in $10$\% increments of the mean X-ray peak achieved. The perspective shows the eruption on the disk with a field of view of 5 stellar radii. Animations (including a limb view) are available in the supplementary material. \textit{Middle:} Light curve of the mean flux ($F_{\rm M}$) relative to the quiescent pre-CME level ($F_{\rm Q}$) on each filter. \textit{Bottom:} Coronal Doppler shift velocity ($\Delta U_{\rm r}$) at the indicated times during the CME-induced flare-like signature. The listed isosurface value in the density contrast ($n(t)/n^{\rm SS}$) is used to identify the perturbation. Selected large-scale magnetic field lines are shown in gray.}
\label{fig_2}
\end{figure*}

\begin{figure*}[!h]
\centering
\includegraphics[trim=0.1cm 0.1cm 0.1cm 0.1cm, clip=true,width=0.995\textwidth]{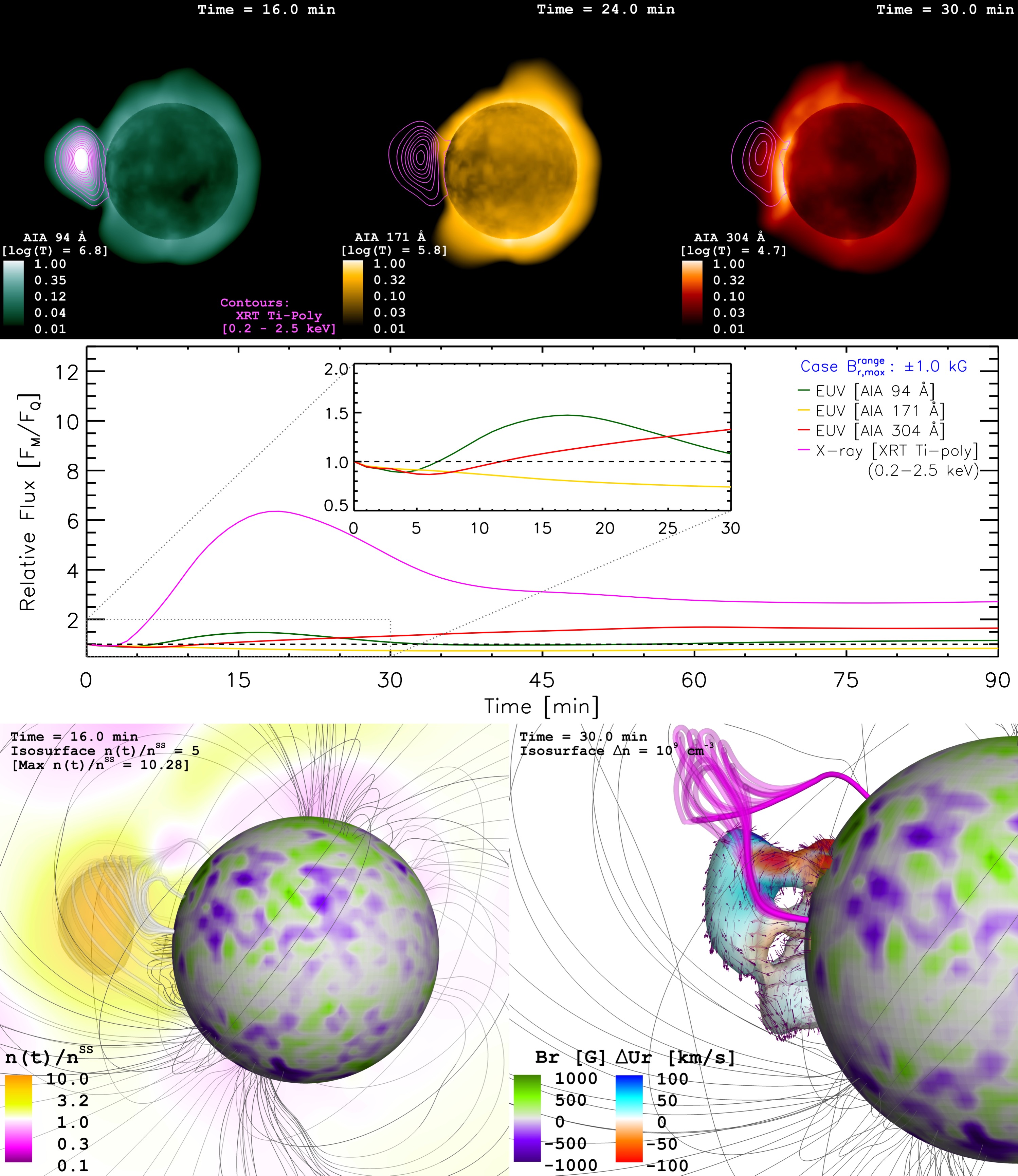}
\caption{Coronal response due to the flux-rope eruption in the $B_{\rm r,\,max}^{\rm range} = \pm1000$ G case. \textit{Top \& Middle:} See caption of~Fig.~\ref{fig_2}. \textit{Bottom:} The left panel contains the coronal density contrast ($n(t)/n^{\rm SS}$), at the time of the 94 \AA~flare peak (16 min after the flux-rope injection). The perturbed material is strongly compressed against the tension exerted by the local magnetic field (white). The right panel shows a later stage in the simulation (30 min of evolution; see top-right panel), focused on the indicated density perturbation, color-coded by Doppler shift velocity ($\Delta U_{\rm r}$). Vectors denote the velocity field of the rising/falling coronal material. The highly twisted magnetic field lines (magenta), precede the escape of the fragmented CME. Selected large-scale magnetic field lines are shown in gray. Animations are available in the supplementary material.}
\label{fig_3}
\end{figure*}

\begin{figure*}[!t]
\centering
\includegraphics[trim=0.1cm 0.1cm 0.1cm 0.1cm, clip=true,width=0.995\textwidth]{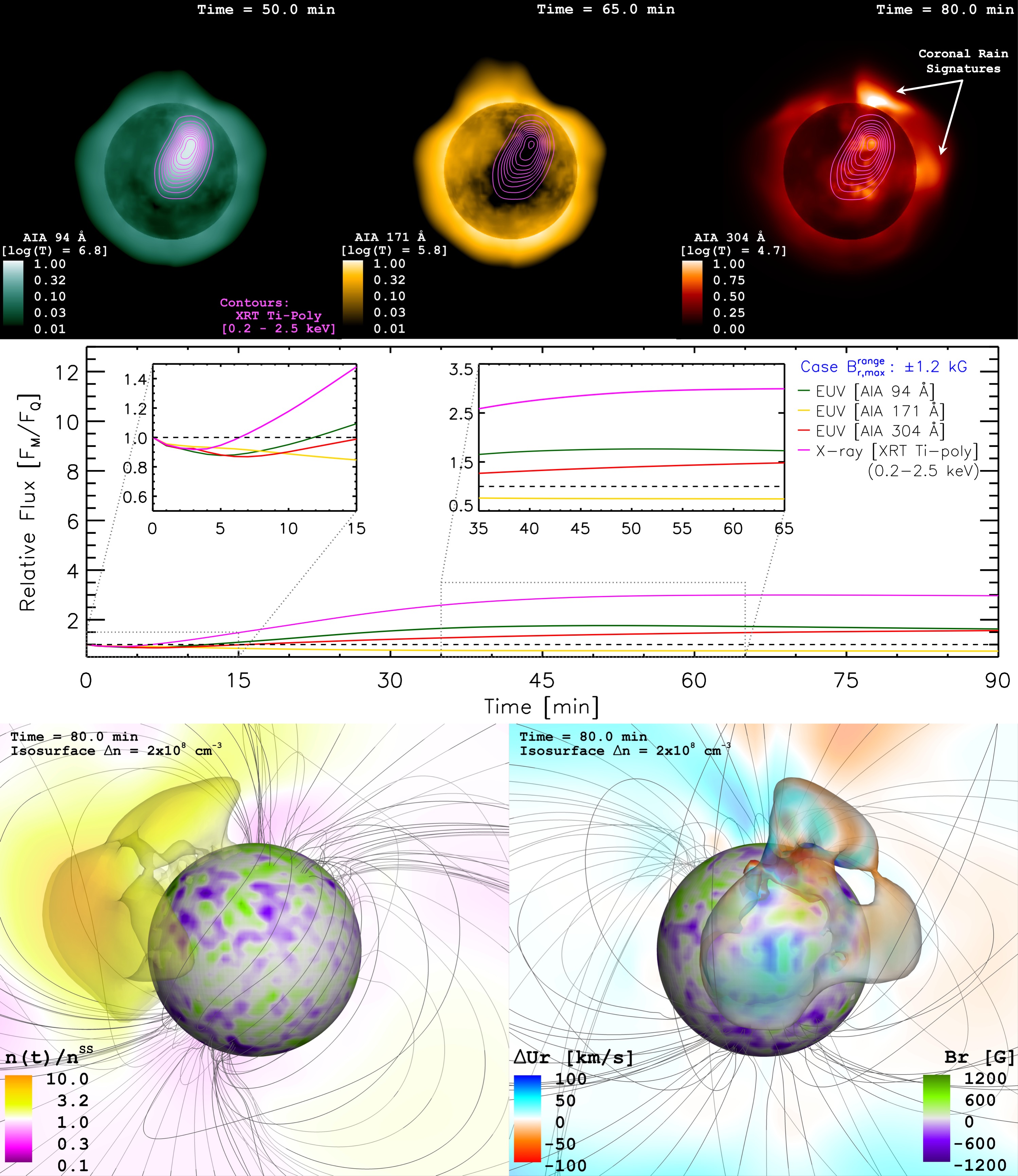}
\caption{Coronal response due to the flux-rope eruption in the $B_{\rm r,\,max}^{\rm range} = \pm1200$ G case. \textit{Top \& Middle:} See caption of Fig.~\ref{fig_2}. \textit{Bottom:} Two 3D views showing a later stage in the simulation (80 min of evolution), focused on the indicated coronal density perturbation, color-coded by density contrast ($n(t)/n^{\rm SS}$, left) and Doppler shift velocity ($\Delta U_{\rm r}$, right). We identify this structure as a cloud of coronal rain, with signatures of the latter appearing in the 304 \AA~channel images (top-right panel). Selected large-scale magnetic field lines are shown in gray. Animations are available in the supplementary material.}
\label{fig_4}
\end{figure*}

\begin{figure}[!t]
\centering
\includegraphics[trim=0.1cm 0.1cm 0.1cm 0.1cm, clip=true,width=0.4083\textwidth]{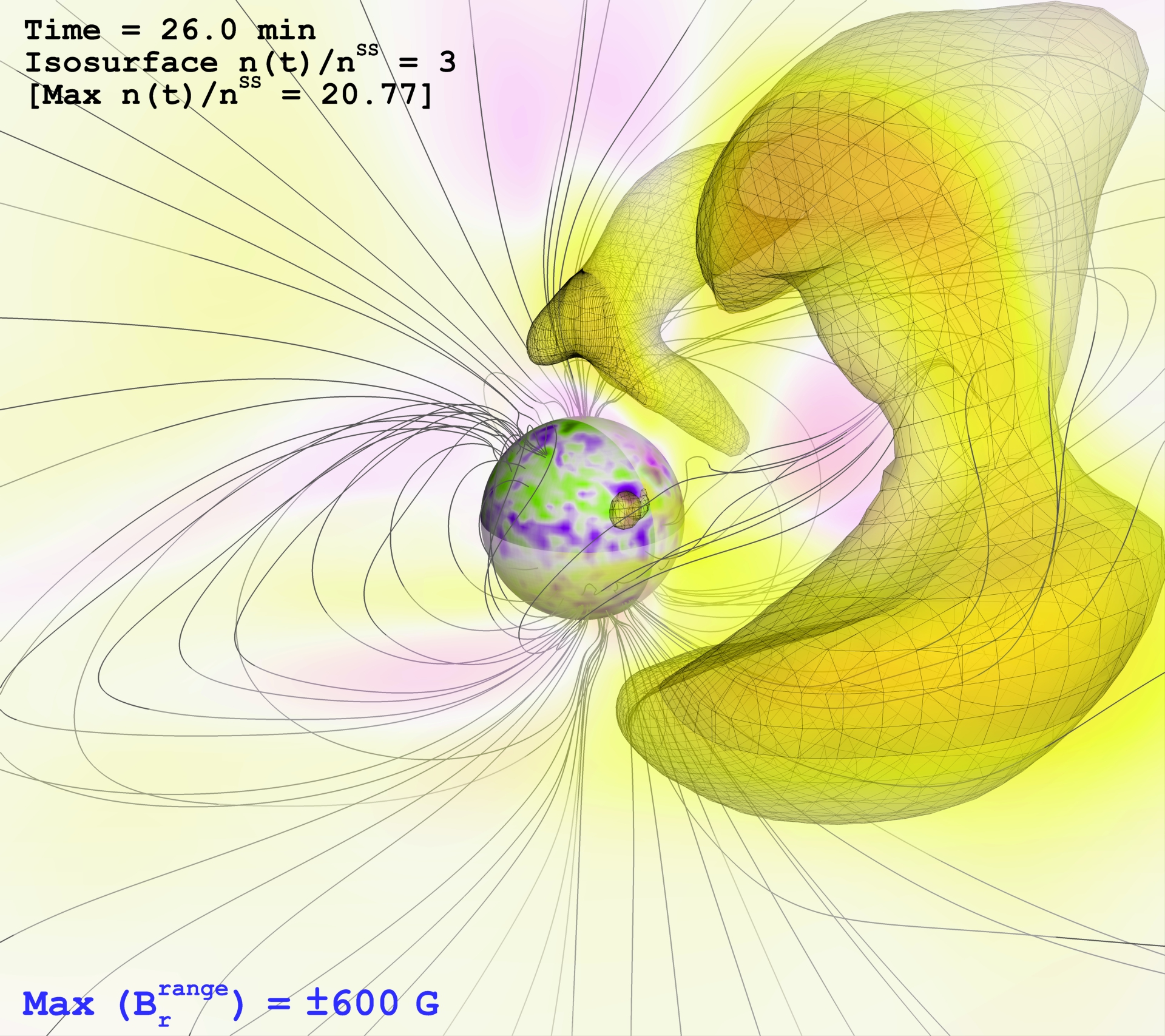}
\includegraphics[trim=0.1cm 0.1cm 0.1cm 0.1cm, clip=true,width=0.4083\textwidth]{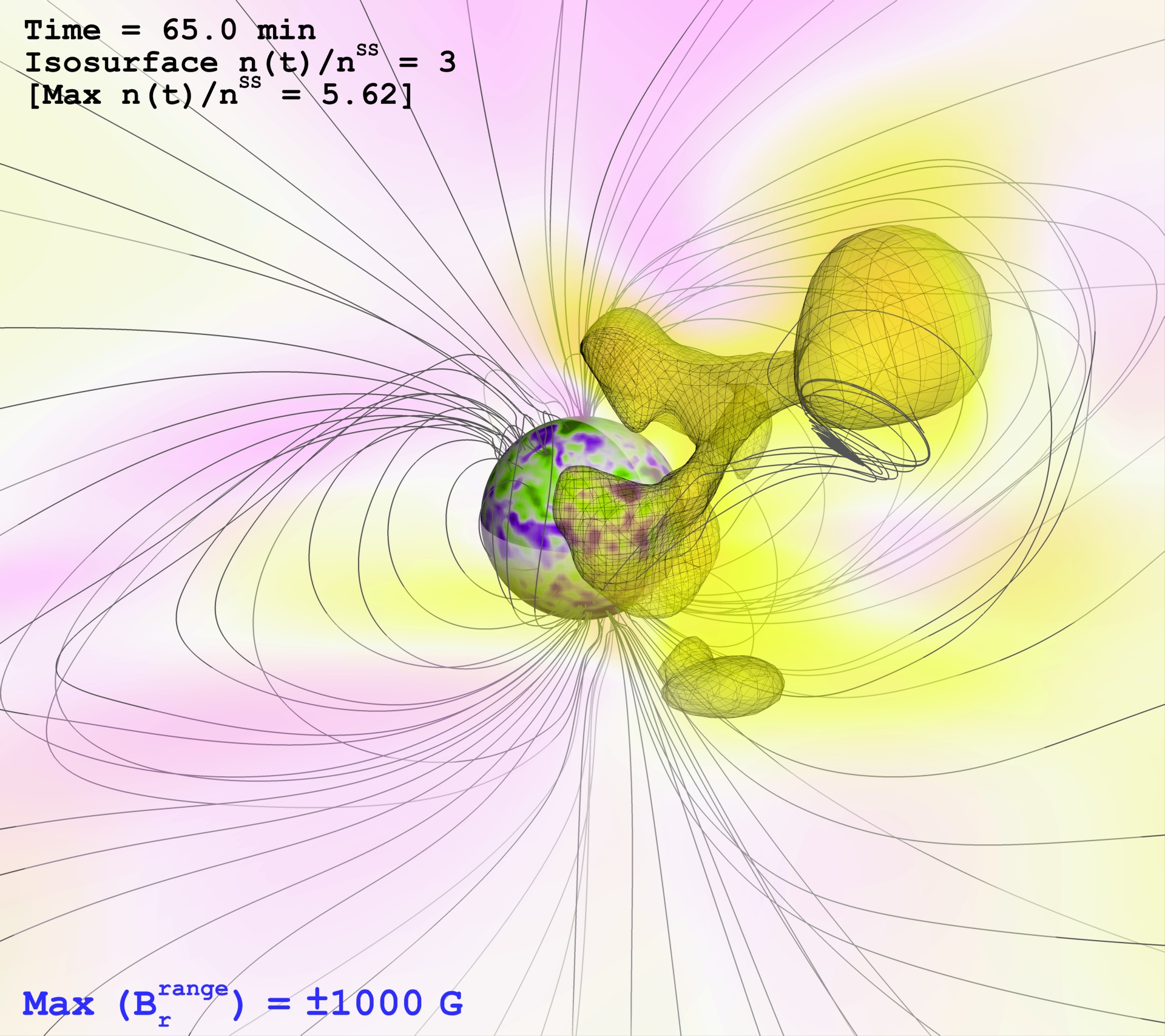}
\includegraphics[trim=0.1cm 0.1cm 0.1cm 0.1cm, clip=true,width=0.4083\textwidth]{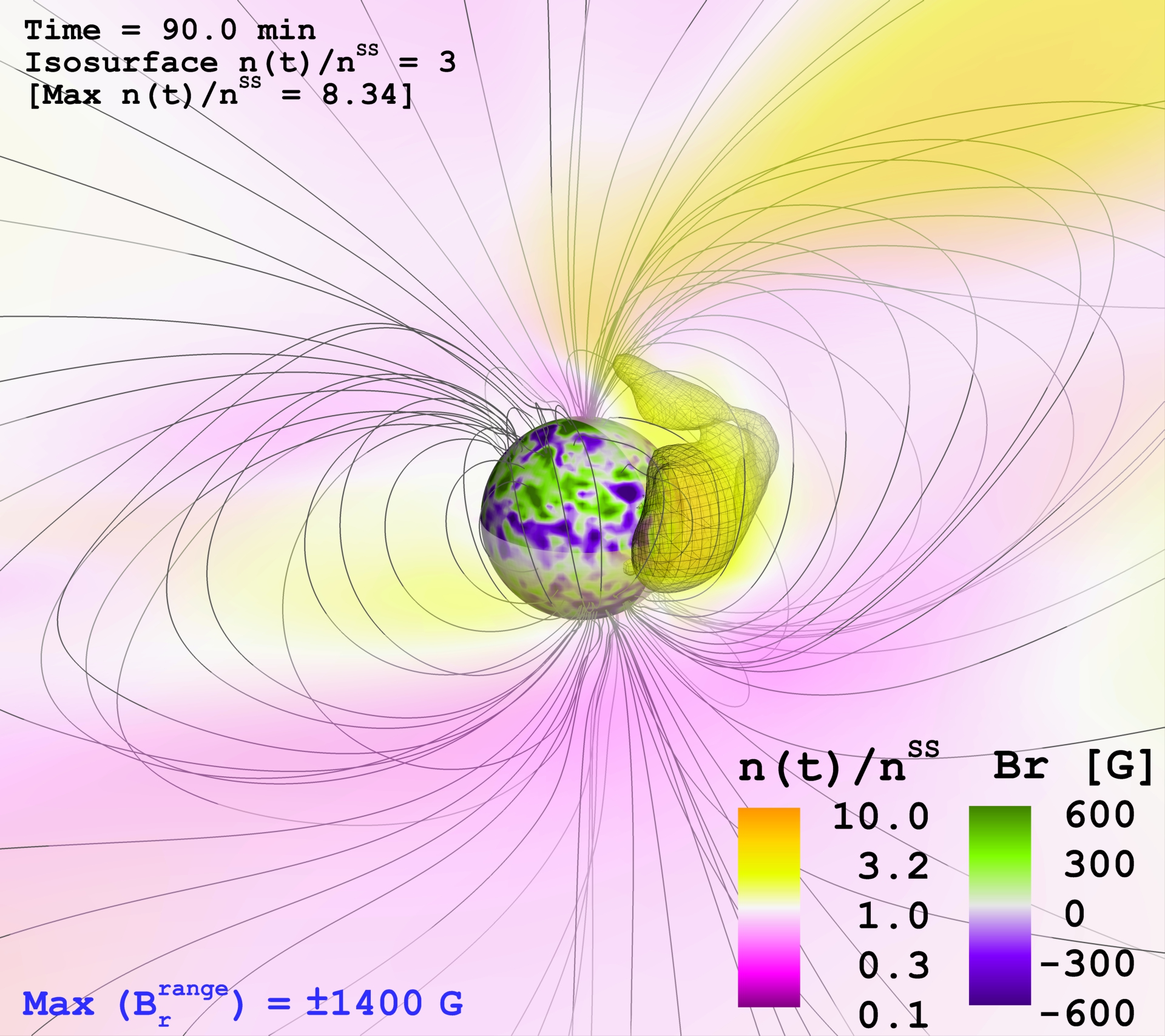}
\caption{Snapshots in the evolution of the weakly (\textit{top}), partially (\textit{middle}), and strongly confined CMEs (\textit{bottom}). Note the time difference between each figure. Colorbars indicate the radial surface field strength (saturated at $\pm600$ G), and the coronal density contrast $n(t)/n^{\rm SS}$. Selected large-scale magnetic field lines are shown in gray. Animations are available in the supplementary material.}
\label{fig_5}
\end{figure}

\noindent Figures \ref{fig_2}, \ref{fig_3}, and \ref{fig_4} contain a summary of the results obtained for three of the magnetic field scalings considered (Sect.~\ref{sec:field}). We include simulated line-of-sight images of the stellar corona, synthesized following response functions of three EUV channels of AIA\footnote[2]{\href{http://aia.lmsal.com/public/instrument.htm}{\textit{Atmospheric Imaging Assembly}}}/SDO\footnote[3]{\href{https://sdo.gsfc.nasa.gov/mission/}{\textit{Solar Dynamics Observatory}}}, as well as contours of the X-ray emission traced by the Ti-Poly filter of XRT\footnote[4]{\href{https://xrt.cfa.harvard.edu/index.php}{\textit{X-Ray Telescope}}}/Hinode\footnote[5]{\href{https://hinode.msfc.nasa.gov/index.html}{\textit{Hinode/Solar-B}}}. Light curves of the mean flux ($F_{\rm M}$) normalized to the pre-CME quiescent level ($F_{\rm Q}$) in each filter, and different 3D visualizations of the events are also included for all cases\footnote[6]{Animations are available in the supplementary material.}. These results are representative of different regimes in the CME confinement spectrum: weak ($B_{\rm r,\,max}^{\rm range} = \pm600$~G, Fig.~\ref{fig_2}), partial ($B_{\rm r,\,max}^{\rm range} = \pm1000$~G, Fig.~\ref{fig_3}), strong ($B_{\rm r,\,max}^{\rm range} = \pm1200$~G,~Fig.~\ref{fig_4}). The status of each eruption is determined by visual inspection of the evolution in density contrast at $n(t)/n^{\rm SS} = 3.0$ (with $n^{\rm SS}$ representing the steady-state pre-CME solution, see Fig.~\ref{fig_5}), and comparing the speed of the perturbation front with the local escape velocity ($v_{\rm esc} = \sqrt{2GM_{*}/H}$, where $H$ indicates the front position and $G$ is the gravitational constant).

\clearpage 
\subsection{Weaker Field Case ($\pm 600$ G)}

\noindent As presented in the top panels of Fig.~\ref{fig_2}, a clear coronal response in all channels is obtained in this case. In the first frames of evolution ($\lesssim$\,$4$ mins), it is possible to observe the collapse of the inserted flux-rope towards the stellar surface, reaching in-falling radial speeds $\gtrsim$\,$400$~km~s$^{-1}$. This collapse is not radial but follows the overarching field configuration, bringing the perturbation equator-wards from the initial launching latitude. During this process the ambient plasma of the underlying canopy is strongly compressed, rapidly increasing the local coronal density and temperature (by factors of $\sim$\,$5-10$), leading to a flare-like signature as registered by the high-temperature sensitive filters (94~\AA~and Ti-Poly). Correspondingly, the mean X-ray flux increases by more than an order of magnitude compared to the quiescent level before the flux-rope insertion (Fig.~\ref{fig_2}, middle panel). A smaller increase is obtained in the 94~\AA~channel, as the average pre-CME emission in this band is higher. The maximum flux in these two filters appears $6-7$~min after the flux-rope eruption. Observations indicate that similar processes could occur in the solar corona on much smaller scales (e.g.,~\citeads{2014ApJ...796...73H}, \citeads{2015Natur.523..437S}), in line with expectations given the relatively weak solar large-scale field. We stress here that our simulation does not include a magnetic reconnection flaring component and only considers the eruption of the flux-rope in the corona.

The collapse continues, reaching lower and cooler layers of the corona, generating an excess in the 304~\AA~channel (visible also in the mean flux; middle panel in Fig.~\ref{fig_2}) until the perturbation \textit{bounces} back due to the strong local magnetic tension. At this stage, surrounding material lifts off with speeds on the order of $50$~km~s$^{-1}$, while the core of the eruption reaches a radial velocity exceeding the local escape value ($\sim$\,$200$~km~s$^{-1}$ for Prox~Cen). Taking into account the initial range of speeds in this region, the resulting coronal Doppler shift average velocities associated with the final eruption would lie in the range of $90 - 150$~km~s$^{-1}$. Snapshots of the Doppler shift evolution can be seen in the bottom panels of Fig.~\ref{fig_2}.

\subsection{Medium Field Cases ($\pm 800$--$1000$ G)}

\noindent The evolution in the $B_{\rm r,\,max}^{\rm range} = \pm1000$~G case (Fig.~\ref{fig_3}) globally follows the previous description, although important differences are clearly visible. The average speeds for the collapse and escape are significantly reduced (by $\sim$\,$40$\% and $30$\%, respectively), which leads to a broader (longer duration) flare-like profile, with a weaker peak that develops at a later time ($16-18$~min after the flux-rope insertion; see Fig.~\ref{fig_3}, middle and bottom-left panel). The stronger stellar field also induces the fragmentation of the perturbation preceding the CME escape (Fig.~\ref{fig_3}, bottom-right panel). A fraction of the eruption remains confined in the corona, displaying rise and fall patterns (with $|\Delta U_{\rm r}|$ up to $100$~km~s$^{-1}$), similar to plasma flows observed during solar coronal rain (\citeads{2012ApJ...745..152A}) or condensations (\citeads{2018ApJ...864L...4L}).

After the escape of the eruption, the corona relaxes to a quasi steady-state with slightly higher levels of emission in some of the analyzed bands (compared to the pre-CME configuration; Figs.~\ref{fig_2} and \ref{fig_3}, middle panel). This was expected due to the permanent modification of the surface magnetic field (required for the flux-rope initiation), and the time-accurate nature of the simulation. Throughout the entire evolution, the mean flux from the 171 \AA~channel remains on average $\sim$\,$15$\% below the quiescent level, with a transient dark sector visible in the corresponding synthetic images (Figs.~\ref{fig_2}, top-centre). This feature clearly resembles the mid-temperature coronal dimmings observed on the Sun, which have been shown to occur consistently during CME-related flaring events (\citeads{2016SoPh..291.1761H}, \citeads{2016ApJ...830...20M}). However, note that the dimming in this band is also visible during fully confined CME events (Fig.~\ref{fig_4}). Given that this flux deficit can occur due to thermodynamic changes in local plasma instead of evacuation and ejecta, this feature alone is probably insufficient to discriminate unambiguously CMEs in stellar observations.

The results for the $B_{\rm r,\,max}^{\rm range} = \pm800$~G scaling (not shown) fall in between the $\pm 600$~G and $\pm 1000$~G scenarios, as expected for a smooth transformation between the regimes of weak and partial CME confinement. This change in global behavior takes place over a relatively large range of surface field values ($\Delta B_{\rm r,\,max}^{\rm range} = 400$ G). 

\subsection{Strong Field Cases ($\pm 1200$--$1400$ G)}

\noindent In the remaining two cases ($B_{\rm r,\,max}^{\rm range} = \pm1200,\pm1400$~G), the erupting flux-rope is not able to escape, so no CME is generated. This indicates that the transition to a strong CME confinement regime is more abrupt in terms of magnetic field increment than the changes between the weaker and moderate field cases, and that the suppression threshold for this particular eruption (Sect.~\ref{sec:cme}) is close to the $\pm1200$--1400~G background field values.

In Fig.~\ref{fig_4} we present the fully confined eruption in the $B_{\rm r,\,max}^{\rm range} = \pm1200$~G case. The response of the corona is more gradual (i.e., less transient), with a much slower rise in the high-energy emission (by factors of $\sim$\,$2-3$), peaking after $\sim$\,$50-60$~min. As mentioned before, the behavior in the 171 \AA~filter shows little difference with respect to its counterpart in the escaping events. In the remaining channels it is possible to observe a small \textit{dip} ($\sim$\,$10-15$\%) in the mean flux during the first few minutes of evolution (Fig.~\ref{fig_4}, middle panel). For a shorter duration, this feature is also visible in the weakly and partially confined CME cases (Figs.~\ref{fig_2} and \ref{fig_3}). Similar pre-flare dips in optical wavelengths have been reported, which, among other possibilities, could be interpreted as possible tracers of CMEs in stars (\citeads{1982ApJ...252L..39G}). For the coronal emission studied here, they are related to the decrement in the emitting volume during the collapse of the flux-rope (common among all cases).

The bottom panels of Fig.~\ref{fig_4} show two views late in the evolution in this event ($80$ min), focused on a density perturbation induced by the failed CME. During its evolution, the disturbance grows until a large fraction of the stellar disk is covered, moving pole- and east-ward with respect to the flux-rope launching site. The structure displays a spatial distribution in Doppler shift velocities within a  $\pm50$~km~s$^{-1}$ range. However, over the course of the simulation, the integrated values for $\Delta U_{\rm r}$ are actually negative (i.e., net in falling material), with a mean time-average of $\sim$\,$-2.5$~km~s$^{-1}$. As such, we identify this structure as a \textit{coronal rain cloud}, which also shows a clear spatial correspondence with signatures visible in lower (and cooler) regions of the corona, captured in the $304$~\AA~images (Fig.~\ref{fig_4}, top-right panel). Interestingly, an extensive observational analysis shows that similar processes might be taking place in the corona of the young solar-analog EK Draconis (see \citeads{2010ApJ...723L..38A}, \citeads{2015AJ....150....7A}).

The system is still dynamically evolving after the $90$~min of evolution captured in our simulation. Nevertheless, a relaxation of the corona into a new quasi-steady state is expected (as observed in the previous cases), with an associated decline in the high-energy emission following standard radiative cooling time-scales of optically thin plasmas (incorporated in the AWSoM; see~\citeads{2014ApJ...782...81V}).

\subsection{Observational Prospects and Conclusions}

\noindent The recondite nature of CMEs on stars contrasts with their importance, both from the perspectives of stellar physics and their impacts on planetary environments. Here we investigated numerically the coronal response during a flux-rope eruption event, under the suppressing influence of magnetic field values expected for moderately active M-dwarfs. Depending on the strength of the stellar field, we obtained a distinctive set of observational signatures of this CME-suppression process:

\begin{itemize}
\item Weakly and partially suppressed CME events lead to increases\,---by factors between $\sim5$ and more than $10$---\,in the integrated X-ray coronal emission ($0.2 - 2.5$ keV) lasting from tens of minutes up to an hour. These brightenings resemble a normal stellar flare, but are instead powered by strong compression of the coronal material rather than by magnetic reconnection.

\item The induced flare-like events display a rapid ($<$$1$~hour) evolution of Doppler shift signals (from red to blue), with velocities up to $200$ km s$^{-1}$, transitioning from hotter ($\log(T) \gtrsim 6.8$) to cooler ($\log(T) \lesssim 6.0$) coronal lines. Lower velocities ($<$$100$ km s$^{-1}$), and longer durations ($\gtrsim 1$ hour), are expected for more strongly confined~eruptions.

\item Fully suppressed CME events result in a gradual brightening of the soft X-ray corona (by factors of $\sim2-3$) over the course of several hours. This emission is redshifted ($<$$-50$~km~s$^{-1}$), indicative of in-falling material (designated here as a coronal rain cloud).
\end{itemize}

None of these phenomena are likely to be observable with current instrumentation, which is generally sensitive to only the most extreme CME candidates and events (\citeads{2019ApJ...877..105M}, \citeads{2019NatAs.tmp..328A}). While velocities revealed in our simulations---90-150~km~s$^{-1}$ in the case of the weakly-suppressed escaping CME---can in principle be discerned by {\it Chandra} (see e.g. \citeads{2019NatAs.tmp..328A}), the problem for observability lies in the short time over which the shifts occur and the low comparative brightness of these less extreme events. The intensity of the induced X-ray emission would be expected to rapidly decline due to its density squared dependence and expansion as the CME is accelerated outward.

However, the coronal response to all of the events simulated here could in principle be observed by more sensitive next generation X-ray missions, such as the {\it ARCUS} and {\it Lynx} concepts (see \citeads{2019BAAS...51c.113D}). Requirements would be a resolving power $\lambda/\Delta\lambda\geq 5000$ to see the relatively small Doppler shifts in the coronal vicinity, and a larger effective area than the {\it Chandra} HETG by at least an order of magnitude, in order to perform X-ray spectroscopy with the required temporal resolution. 

\acknowledgments
\noindent We would like to thank the referee for constructive feedback. J.D.A.G. was supported by Chandra GO5-16021X and HST GO-15326 grants. J.J.D. was funded by NASA contract NAS8-03060 to the Chandra X-ray Center and thanks the director, Belinda Wilkes, for continuing advice and support. S.P.M. and O.C. were supported by NASA Living with a Star grant number NNX16AC11G. This work was carried out using the SWMF/BATSRUS tools developed at The University of Michigan Center for Space Environment Modeling (CSEM) and made available through the NASA Community Coordinated Modeling Center (CCMC). Resources supporting this work were provided by the NASA High-End Computing (HEC) Program through the NASA Advanced Supercomputing (NAS) Division at Ames Research Center. Simulations were performed on NASA's Pleiades cluster under award SMD-17-1330. This work used the Extreme Science and Engineering Discovery Environment (XSEDE), which is supported by National Science Foundation grant number ACI-1548562. This work used XSEDE Comet at the San Diego Supercomputer Center (SDSC) through allocation TG-AST170044.

%

\facilities{NASA Pleiades Supercomputer, XSEDE \citepads{2014CSE....16e..62T} Comet/SDSC}


\software{SWMF \citepads{2018LRSP...15....4G}}

\end{document}